\documentclass{article}

\usepackage{arxiv}

\usepackage[utf8]{inputenc} 
\usepackage[T1]{fontenc}    
\usepackage[hidelinks]{hyperref}
\usepackage{url}            
\usepackage{booktabs}       
\usepackage{amsfonts}       
\usepackage{nicefrac}       
\usepackage{microtype}      
\usepackage{lipsum}
\usepackage{graphicx}
\usepackage{bm}

\usepackage{amsmath,mathrsfs,enumerate,multirow,appendix}
\usepackage{longtable, caption}
\usepackage[super,sort&compress]{natbib}
\bibpunct{}{}{,}{s}{}{}

\usepackage{tikz}
\usetikzlibrary{decorations.pathreplacing, arrows.meta}

\makeatletter
\renewcommand\@biblabel[1]{#1.}
\makeatother

\graphicspath{ {./images/} }

\title{Testing Prioritized Composite Endpoint with Multiple Follow-up Time Examinations}

\author{
 Yunhan Mou \\
  Department of Biostatistics,\\
  Yale School of Public Health\\
  New Haven, Connecticut, USA \\
  \texttt{yunhan.mou@yale.edu} \\
   \And
 Haitao Pan \\
  Department of Biostatistics,\\
  St. Jude Children's Research Hospital\\
  Memphis, Tennessee, USA \\
  \texttt{haitao.pan@stjude.org} \\
  \And
 Yu Jiang \\
  Division of Epidemiology, Biostatistics \\ and Environmental Health,\\
  School of Public Health, University of Memphis\\
  Memphis, Tennessee, USA\\
  \texttt{yjiang4@memphis.edu} \\
  \And
  Yuan Huang\thanks{Corresponding author}\\
    Department of Biostatistics,\\
    Yale School of Public Health\\
    New Haven, Connecticut, USA; \\
    \texttt{yuan.huang@yale.edu}\\
}

\begin{document}
\maketitle
\vspace{-0.1in}
\begin{abstract}
Composite endpoints are widely used in cardiovascular clinical trials. In recent years, hierarchical composite endpoints—particularly the win ratio approach and its predecessor, the Finkelstein-Schoenfeld (FS) test, also known as the unmatched win ratio test—have gained popularity.  These methods involve comparing individuals across multiple endpoints, ranked by priority, with mortality typically assigned the highest priority in many applications. However, these methods have not accounted for varying treatment effects, known as non-constant hazards over time in the context of survival analysis. To address this limitation, we propose an adaptation of the FS test that incorporates progressive follow-up time, which we will refer to as ProFS.  This proposed test can jointly evaluate treatment effects at various follow-up time points by incorporating the maximum of several FS test statistics calculated at those specific times. Moreover, ProFS also supports clinical trials with group sequential monitoring strategies, providing flexibility in trial design. As demonstrated through extensive simulations, ProFS offers increased statistical power in scenarios where the treatment effect is mainly in the short term or when the second (non-fatal) layer might be concealed by a lack of effect or weak effect on the top (fatal) layer. We also apply  ProFS to the SPRINT clinical trial, illustrating how our proposed method improves the performance of FS.
\end{abstract}

\keywords{Composite Endpoint \and Win Statistics\and
            Time-varying Effects \and Generalized Pairwise Comparisons 
            \and Survival Time }

\section{Introduction}

Composite endpoints are frequently employed to measure treatment effects in cardiovascular trials. A commonly used endpoint is time to first occurrence of death or hospitalization, which combines a fatal event (death) with a non-fatal event (hospitalization). However, such a time-to-first-event endpoint assumes equal importance for all events, ignoring the fact that death is clinically far more serious than hospitalization. To address this issue, Pocock et al. \cite{pocock2012win} introduced the win ratio (abbreviated as WR for either the method or the win ratio measure) method, which performs pairwise comparisons between patients using a hierarchical structure, prioritizing the time-to-death endpoint in the comparison between each pair of patients. By doing so, the WR method aligns the analysis with clinical priorities and ensures that more serious events receive appropriate attention in evaluating treatment effects. The core testing strategy for the unmatched WR \cite{pocock2012win} followed the Finkelstein-Schoenfeld (FS) test \cite{finkelstein1999combining}. 
In recent years, WR has gained increasing popularity and has been adopted in a range of clinical trials, including EMPULSE (registration number in ClinicalTrials.gov: NCT0415775), DAPA-HF (NCT03036124), VIP-ACS (NCT04001504), and CanCovDia (NCT04510493). Notably, the U.S. Food and Drug Administration (FDA) recognized the method in its 2022 guidance on multiple endpoints \cite{FDA2022}, reflecting regulatory support for its broader adoption. 
Beyond cardiovascular trials, WR has shown promise in complex clinical settings where multiple domains of benefit need to be jointly evaluated. For instance, a post hoc application of WR to the COMET trial (NCT02782741) in late-onset Pompe disease demonstrated that WR could effectively integrate respiratory and mobility outcomes in a hierarchical fashion, yielding a win ratio of 2.37 in favor of the experimental enzyme therapy \cite{boentert2024post}. This example illustrates WR’s suitability for rare disease trials, where small sample sizes and heterogeneous outcomes often limit traditional analytic strategies. Furthermore, the method has been increasingly explored in oncology, where traditional composite endpoints often fail to capture the clinical tradeoffs between survival and quality of life or functional outcomes. As highlighted by Pocock et al. \cite{pocock2012win}, WR’s ability to incorporate different types of measures makes it well-suited for trials with multiple, competing endpoints. These advances underscore WR’s flexibility and its increasing value in improving interpretability and statistical robustness across therapeutic areas.

Beyond the expanding clinical applications, WR has also inspired a growing body of methodological research. To address trial designs involving stratification, Dong et al. \cite{dong2018stratified} and Gasparyan et al. \cite{gasparyan2021adjusted} introduced the stratified win ratio, which allows comparisons within strata and aggregates evidence across them. A number of statistical inference techniques have been developed for the win ratio, including methods for constructing confidence intervals and formal hypothesis testing \cite{luo2015alternative, bebu2016large, dong2016generalized, mao2019alternative}. The challenge of censoring has also received attention. Oakes \cite{oakes2016win} proposed an integral form of WR to account for censoring, while Dong et al. \cite{dong2020inverse, dong2021adjusting} developed inverse-probability-of-censoring weighting (IPCW) approaches to mitigate bias from censored data. Mao \cite{mao2024defining} further clarified the estimand underlying WR and emphasized its dependence on the chosen time frame, highlighting the importance of aligning analytical strategies with clinically meaningful durations. In the border family of methods, win statistics (or generalized pairwise comparisons), to which FS and WR both belong, there are more variations that share a similar concept of prioritizing endpoints. Related methods include the generalized pairwise comparisons (or net benefit approach) \cite{buyse2010generalized}, win-loss statistics \cite{luo2017weighted}, win odds \cite{brunner2021win}, win probability \cite{gasparyan2021adjusted}, and the event-specific win ratio \cite{yang2021eventct,yang2022eventsim}. A comprehensive overview of this family is available in Verbeeck et al. \cite{verbeeck2023generalized}, and issues specific to censoring are discussed by P{\'e}ron et al. \cite{peron2018extension} and Deltuvaite-Thomas et al. \cite{deltuvaite2023generalized}. Additionally, regression-based win function modeling has been explored to evaluate covariate effects \cite{mao2021class, wang2022stratified, song2023win}.

Despite these advances, one key limitation persists: neither the initial FS nor the subsequent win statistics are designed to accommodate poential short-term treatment effects. In many clinical settings, treatment effects may differ between short-term and long-term follow-up periods \cite{jatoi2016time}. For example, in trials comparing endovascular repair versus open repair for abdominal aortic aneurysm, the survival benefit of endovascular repair is more evident in the short term than the long term \cite{lederle2009outcomes, lederle2012long}. In such cases, using a fixed-length follow-up in the FS test may obscure early benefits or fail to detect treatment differences altogether. This limitation stems not from the testing strategy per se, but from the null hypothesis formulation, which implicitly assumes a constant treatment effect over time.
To address this, a more flexible analytical framework that jointly assesses short- and long-term effects may provide a more accurate reflection of treatment benefit and improve sensitivity in detecting potential short-term effects.
Acknowledging the strength of FS in combining multiple endpoints and the need for jointly testing treatment effects at different lengths of follow-up time (i.e., short and long terms), we introduce the Progressive Follow-up Time FS test (ProFS). This method constructs multiple FS test statistics based on data observed at different pre-specified follow-up time points, referred to as examinations.
For use in trial settings, examination times can be scheduled to coincide with routine clinical assessments when endpoint collection is tied to upcoming visits. In other cases, for continuously monitored information, such as death, hospitalization, the determination of  examination times may be more flexible. 
In particular, we explore a  quantile-based rule to generate examination time points in a principled and reproducible way. 
Rather than analyzing each time point separately, ProFS uses U-statistic theory to form a joint test statistic under the asymptotic multivariate normal distribution of these FS scores, thereby assessing whether the maximum difference across all examinations is statistically significant. Under the null hypothesis, ProFS assumes no treatment difference at any of the pre-specified follow-up time points, offering a principled framework for testing over time while controlling type I error inflation—a common concern when performing repeated tests on the same patients.
Moreover, ProFS offers additional advantages in those hierarchical settings where the conventional FS test may underperform when treatment effects are concentrated in lower-priority endpoints. 
We demonstrate the utility of ProFS through extensive simulation studies that vary the treatment effect size, correlation structure between endpoints, and follow-up duration. In addition, we apply ProFS to the Systolic Blood Pressure Intervention Trial (SPRINT) (NCT01206062) to illustrate its real-world performance. Finally, we extend ProFS to accommodate group sequential trial designs, enabling broader application in interim analysis settings.

The remainder of this paper is structured as follows. Section 2 introduces the proposed ProFS methodology, including its extension to accommodate clinical trials with group sequential designs. Section 3 presents results from simulation studies designed to evaluate the performance of ProFS under various scenarios. Section 4 applies the proposed method to the Systolic Blood Pressure Intervention Trial (SPRINT) to illustrate its practical utility. Finally, Section 5 concludes with a discussion of key findings and future research directions.

\section{Method}\label{method}

In this section, we first review the standard FS and then propose our ProFS. For simplicity, we consider a clinical trial setting with two endpoints of interest, time to death and time to hospitalization.
Suppose there are $N$ participants, out of which $M$ are in the treatment group. For the $i$-th participant,  $T_i=0$ if the  participant is in the control group and $T_i=1$ if the  participant is in the treatment group ($M=\sum_{i=1}^{N}{T_i}$ ).
Let $D_i$ and $C_{Di}$ be the observed time to death and censoring indicator, respectively, such that  $C_{Di}=0$ if the death event is observed. Similarly, let $H_i$ and $C_{Hi}$ be the observed time to hospitalization and its censoring indicator, respectively. 
The primary interest is to test the difference between treatment and control groups, where longer time to death and time to hospitalization is preferred.

\subsection{Standard FS Test}
FS is based on pair comparisons among all participants. For each pair of participants $i$ and $j$, a score $u_{ij}$ is assigned to reflect whether participant $i$ has a more favorable performance than $j$ such that $u_{ij}$ is 1 if $i$ outperforms $j$ (win), $-1$ if $j$ outperforms $i$ (loss), and $0$ if the comparison is uninformative or indeterminate (tie).
To determine $u_{ij}$, comparisons will be made across multiple endpoints in a hierarchical manner. We first examine the time-to-death information and determine if one lives longer than the other. If $i$ and $j$ have the same time to death (or if a tie arises due to censoring), the time-to-hospitalization endpoint is examined to determine whether one participant has a longer time to hospitalization than the other. If there is still no determinate result with either the same time to hospitalization or censoring, a tie will be concluded for the comparison between $i$ and $j$. After performing pairwise comparisons with all other participants ($j \ne i$), the score for the $i$-th participant is computed as $U_i = \sum_{j \ne i} u_{ij}$. The test is then constructed based on $Z = \sum_{i=1}^{N}U_i T_i$. Under the null hypothesis, where there is no difference between treatment and control groups, $Z$ follows a normal distribution with mean zero and estimated variance in a closed form as $\widehat{\text{Var}}(Z) = \frac{M(N-M)}{N(N-1)}\left( \sum_{i=1}^{N} U_i^2\right)$
asymptotically. 

\subsection{Progressive Follow-up Time FS Test}
Taking the potential shorter-term treatment effects into account, we propose ProFS. The key idea of ProFS is to compare treatment and control at several different time points simultaneously. 
Suppose the total scheduled follow-up time is $S$, and that $p$ examination time points ${S^{(1)}, ..., S^{(p)}}$ are pre-specified in the protocol, typically aligned with key clinical assessments. For the $k$-th examination at time $S^{(k)} \leq S$, let  $D_i^{(k)}$ and $H_i^{(k)}$ denote the observed time to death and hospitalization, respectively, with $ C_{Di}^{(k)}$ and $C_{Hi}^{(k)}$ representing their respective censoring indicators. 
Accordingly, FS statistic $Z^{(k)}$ and its variance $\text{Var}(Z^{(k)})$ can be calculated for testing 
$H_0^{(k)}$: there is no difference between treatment and control groups at examination time $S^{(k)}$.
 
Combining all examinations, the primary interest becomes testing the joint null hypothesis, $H_0 = \bigcap_k H_0^{(k)}$: there is no difference between the treatment and control groups at any examination time of $S^{(1)},...,S^{(p)}$. According to the multivariate U-statistics theory \citep{lehmann1963robust}, under the null hypothesis, $\bm{Z}=(Z^{(1)},Z^{(2)},...,Z^{(p)})^\top$ is a limiting $p$-variate normal distribution with mean zero and covariance matrix $\bm{\Sigma}$. For $\bm{\Sigma}$, the closed-form estimation is 
\begin{equation}\label{cov mat}
	\hat{\bm{\Sigma}}_{p\times p}= (\hat{\sigma}_{k_1 k_2})=
	  	\left\{
    	\begin{array}{ll}
    	\frac{M(N-M)}{N(N-1)}\left( \sum_i {U_i^{(k_1)}}^2\right) 		& ~~~k_1=k_2,\\
    	\frac{M(N-M)}{N(N-1)}\left( \sum_i U_i^{(k_1)} U_i^{(k_2)} \right)    & ~~~k_1\neq k_2.
    	\end{array} \right. 
\end{equation}

The joint null hypothesis $H_0$ can be tested with the maximum test. This max-type approach is particularly suitable for detecting a signal at any time point, offering robustness against diluted effects in later follow-up periods.
Let
\begin{equation}
    Z_{\mathrm{MAX}} = \max (|R_1|,|R_2|,...,|R_p|),
\end{equation}
where $R_k = Z^{(k)}/\sqrt{\text{Var}(Z^{(k)})}\ (k=1,2,...,p)$ are the standardized test statistics calculated at each examination. For $Z_{\mathrm{MAX}}$ and $z \geq 0$, under the null hypothesis, it holds asymptotically that 
\begin{eqnarray}\label{Z_max<z}
    \mathbb{P}( Z_{\mathrm{MAX}} \leq z) && 
                = \mathbb{P}({-z \leq R_k \leq z, \forall k=1,2,...,p}) \\
                && = \int_{r_1\in[-z,z]} ...\int_{r_p\in[-z,z]} \varphi_{R_1,...,R_p}(r_1,...,r_p) \mathrm{d}r_p,...,r_1,
\end{eqnarray}
where $\varphi_{R_1,...,R_p}(r_1,...,r_p)$ is the probability density function of the limiting joint distribution of $(R_1,...,R_p)^\top$, a $p$-variate normal distribution with mean $\bm 0$ and covariance matrix $\bm{\Omega}$. The estimated $\bm{\Omega}$ is the correlation matrix corresponding to $\hat{ \bm{\Sigma}}$:
\begin{equation}\label{cor mat}
	\hat{\bm{\Omega}}_{p\times p}= (\hat{\omega}_{k_1 k_2})=
	  	\left\{
    	\begin{array}{ll}
    	1 		&~~~k_1=k_2,\\
    	\frac{\sum_i U_i^{(k_1)} U_i^{(k_2)}}{\sqrt{\left( \sum_i {U_i^{(k_1)}}^{2}\right) \left( \sum_i {U_i^{(k_2)}}^{2}\right) }}    & ~~~k_1\neq k_2.
    	\end{array} \right. 
\end{equation}
This probability can be numerically computed \cite{genz1993comparison,Genz1992numericalcompu} with the R package ``mvtnorm'' \cite{genz2009computation}.
The p-value of the maximum test is then given by $P = 1 - \mathbb{P}(Z_{\mathrm{MAX}} \leq \hat Z_{\mathrm{MAX}})$, where $\hat Z_{\mathrm{MAX}}$ is the observed value of  $Z_{\mathrm{MAX}}$. With the maximum test, the treatment effects at $p$ examinations are jointly tested using a single test statistic. To explicitly reflect that the results depend on the chosen examination schedule, we denote the procedure as $\text{ProFS}(S^{(1)},...,S^{(p)})$, which emphasizes its dependence on the timing of scheduled assessments.

\subsection{Selecting Examination Times via Quantile Values}

In this section, we introduce a pre-specified approach to determine the examination times, $S_1,...,S_p$, when there is no sufficient clinical information available. In practice, additional considerations should be taken into account, such as clinical rationale, scheduled visit windows, or cumulative
information fractions, to ensure regulatory acceptability and interpretability. For example, endpoint information may depend on scheduled visits, or clinical knowledge may inform when examination times should be arranged to align with the expected onset of treatment effects.
When such clinical knowledge is lacking and endpoint information does not depend on clinical visits, the approach introduced here provides a framework for predefining examination times.

Suppose there are $p$ examinations. We consider the framework that places $p$ equally spaced time points between an early follow-up threshold and the full study duration $S$. The examination times $(S^{(1)},...,S^{(p)})$ are specified as follows:
\begin{equation}\label{quantile time}
	(S^{(1)},...,S^{(p)})=
	  	\left\{
    	\begin{array}{ll}
    	(\frac{1}{p}S,\frac{2}{p}S,...,S) 		& ~~~S/p \geq S_{\text{inf}},\\
     (S_{\text{inf}},S_{\text{inf}}+\frac{1}{p-1}(S-S_{\text{inf}}),...,S)
        & ~~~S/p < S_{\text{inf}}.
    	\end{array} \right. 
\end{equation}
Here $S_{\text{inf}}$ represents the earliest time to be considered for examination, which can be pre-specified based on clinical indications or the estimated time required for a certain number of events to occur following the study design specifications on hazard rates and recruitment speed, such that the determination is not relied on the observed data. 

The choice of $p$ governs the temporal resolution of ProFS. Following common practice in exploratory data analysis, we recommend $p = 4$  as a default.
This default strikes a balance between temporal granularity and statistical interpretability, akin to the common use of quartiles in descriptive analyses.
We will further explore the influence of $p$ with simulation in Section \ref{emp com}. 
Intuitively, adding an examination may increase power if the additional information accentuates differences between groups enough to offset the increased penalty for controlling type I errors. On one hand, the observed $\max(\hat Z_{\text{MAX}}, \hat R_{p+1})$ is non-decreasing with an added examination at $p+1$. On the other hand,   $\mathbb{P}(\max(Z_{\text{MAX}}, R_{p+1}) \leq z) = \mathbb{P}(Z_{\text{MAX}} \leq z  \text{ and }  -z \leq R_{p+1} \leq z )\leq \mathbb{P}(Z_{\text{MAX}} \leq z)$. Hence the p-value,  $1-\mathbb{P}(\max(Z_{\text{MAX}}, R_{p+1}) \leq z)$, is also non-decreasing with the added examination. 
It is advisable to select an appropriate $p$ by carefully considering the study's designed follow-up length, the mechanism of events, the conditions of the target participant population, and other relevant factors. Clinical trials with longer follow-up lengths or more frequent changes in patients' conditions may consider a larger number of examinations. 
It is important to pre-specify  $p$ and examining times, as making changes after conducting the test may compromise control of the type I error rate. 

Under the equal time segmentation framework, determining examination times simplifies to selecting $p$ when no requirement is imposed on $S_{\text{inf}}$. When additional considerations such as clinical rationale or endpoint availability are relevant, examination times should be chosen to account for those factors. Nonetheless, the proposed framework can still offer preliminary guidance in the design stage. In our simulation, we adopt this framework as a flexible and reproducible approach for generating examination times, enabling us to evaluate the operating characteristics of ProFS under different follow-up granularities and to examine how robustness and power vary across settings.

\subsection{Adaptation to Group Sequential Design}

Group sequential design is a type of adaptive design that provides flexibility and enables early stopping based on interim results.
Here, we focus primarily on stopping for efficacy and derive a method to compute the corresponding boundaries for the adjusted nominal levels \citep{pocock1977group, o1979multiple}.

Let $Q$ be the number of scheduled interim looks, with each interim analysis including an incremental $2l$ participants, equally allocated between the treatment and control groups. For these $2l$ participants, follow-up until primary evaluation should be completed and the primary endpoints are available for assessment.
Define the stopping boundaries $b_1,...,b_Q$ as chosen with respect to the pre-specified probabilities of efficacy stops at these looks, $\tau_1,...,\tau_Q$, which is usually an increasing sequence with $\tau_Q=0.05$.  At the $q$-th look, ProFS maximum test statistic $Z_{\mathrm{MAX}}^{q}$ is obtained. The trial is stopped early for superiority if $Z_{\mathrm{MAX}}^{1} > b_1$ (i.e., early stop at the first look) or $Z_{\mathrm{MAX}}^{1} \leq b_1, Z_{\mathrm{MAX}}^{2} > b_2$ (i.e., early stop at the second look) or $Z_{\mathrm{MAX}}^{1} \leq b_1, Z_{\mathrm{MAX}}^{2} \leq b_2, Z_{\mathrm{MAX}}^{3} > b_3$ (i.e., early stop at the third stop)  and so on. If none of these conditions are met, the final conclusion is drawn at the end of the study using $Z_{\mathrm{MAX}}^{Q}$.

\begin{figure}[h!]
	\centering
\begin{tikzpicture}[>=Stealth, node distance=1.2cm]

\node at (-4,1) {\textbf{At Look 1}};
\node at (-4,0) {$i \in \{1,2,\ldots,2l\}$};

\draw[line width=0.8pt] (-1,0) -- (2,0);

\draw[line width=0.3pt] (-1,0) -- (-1,0.2);
\draw[line width=0.3pt] (0.5,0) -- (0.5,0.2);\node at (0.5,0.5) {$\frac{1}{2}\,\mathrm{FU}$};
\draw[line width=0.3pt] (2,0) -- (2,0.2);\node at (2,0.5) {$\mathrm{FU}$};

\draw[dashed,->] (0.5,0) -- (0.5,-0.6) node[midway,right] {} node[below] {$Z^{1(1)}$};
\draw[dashed,->] (2,0) -- (2,-0.6) node[midway,right] {} node[below] {$Z^{1(2)}$};

\draw[->, thick] (2.5,-0.5) -- (4,-0.5) node[midway,above] {} node[right] {$Z^1_{\mathrm{MAX}}$};
\end{tikzpicture}

\vspace{1cm} 


\begin{tikzpicture}[>=Stealth, node distance=1.2cm]
\node at (-4,1) {\textbf{At Look 2}};
\node at (-4,-0) {$i \in \{1,2,\ldots,2l\}$};
\node at (-4,-1.8) {$i \in \{2l+1,\ldots,4l\}$};

\draw[line width=0.8pt] (-1,0) -- (2,0);
\draw[line width=0.3pt] (-1,0) -- (-1,0.2);
\draw[line width=0.3pt] (0.5,0) -- (0.5,0.2);\node at (0.5,0.5) {$\frac{1}{2}\,\mathrm{FU}$};
\draw[line width=0.3pt] (2,0) -- (2,0.2);\node at (2,0.5) {$\mathrm{FU}$};

\draw[dashed,->] (0.5,0) -- (0.5,-0.6);
\draw[dashed,->] (2,0) -- (2,-0.6);

\node at (0.5,-0.85)  {$Z^{2(1)}$};
\node at (2,-0.85)  {$Z^{2(2)}$};
\draw[line width=0.8pt] (-1,-1.8) -- (2,-1.8);
\draw[line width=0.3pt] (-1,-1.8) -- (-1,-1.6);
\draw[line width=0.3pt] (0.5,-1.8) -- (0.5,-2.0);\node at (0.5,-2.4) {$\frac{1}{2}\,\mathrm{FU}$};
\draw[line width=0.3pt] (2,-1.8) -- (2,-2.0);\node at (2,-2.4) {$\mathrm{FU}$};

\draw[dashed,->] (0.5,-1.8) -- (0.5,-1.2);
\draw[dashed,->] (2,-1.8) -- (2,-1.2);

\draw[->, thick] (2.5,-0.85) -- (4,-0.85) node[midway,above] {} node[right] {$Z^2_{\mathrm{MAX}}$};

\end{tikzpicture}
\caption{Structure of ProFS test statistics in group sequential trials with $\text{ProFS}(0.5S,S)$ and $Q=2$ interim looks.}
\label{fig:SQM illustrate}
\end{figure}
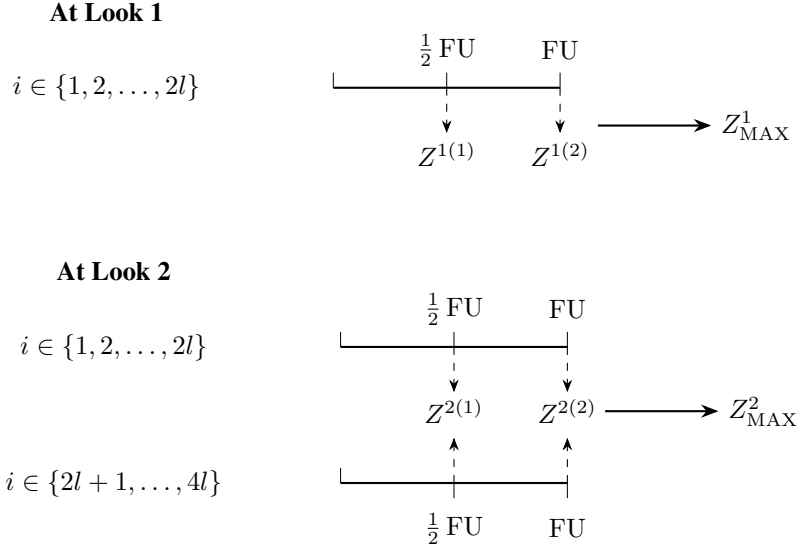


To illustrate the structure of the test statistics at interim looks, a simple example with $\text{ProFS}(0.5S,S)$ and $Q=2$ is shown in Figure \ref{fig:SQM illustrate}. In general, $\text{ProFS}(S_1,...,S_p)$ is employed and $Q$ looks are scheduled. Considering each interim look as a stratum, we have
\begin{eqnarray}
   &Z^{q(k)} &= \sum_{j=1}^{q} \sum_{i=2l(j-1)+1}^{2lj} U_i^{q(k)}T_i, \label{equ: SQM1 Z1} \\
   &\widehat {\text{Var}} (Z^{q(k)}) &= \frac{l^2}{2l(2l-1)} \sum_{j=1}^{q} \sum_{i=2l(j-1)+1}^{2lj} (U_i^{q(k)})^2,  \label{equ: SQM1 VAR Z1}\\
   &\widehat {\text{Cov}} (Z^{q(k_1)},Z^{q(k_2)}) &= \frac{l^2}{2l(2l-1)}\sum_{j=1}^{q} \sum_{i=2l(j-1)+1}^{2lj} U_i^{q(k_1)} U_i^{q(k_2)},~~~k_1\neq k_2 \label{equ: SQM1 COV Z1}
\end{eqnarray}
where $Z^{q(k)}$ stands for the test statistic obtained from the $k$-th examination time point at the $q$-th look. Finally, $Z_{\mathrm{MAX}}^{q} =  \max\{ Z^{q(1)}, Z^{q(2)}, ..., Z^{q(p)} \}$.

We adopt simulation to determine the boundaries. The general idea is that the boundary at each interim analysis $(q \in \{1, \ldots, Q\})$ is determined as the $V(1-\tau_q)$-th smallest value among a set of $V$ elements, which consist of the observed test statistic and $V-1$ values simulated from the null distribution.
Specifically, let $\bm Z^q = (Z^{q(1)},Z^{q(2)},\ldots,Z^{q(p)})$ be the vector of observed test statistics at the $q$-th look. Under the null, $\bm Z^q$ follows asymptotic normal distribution $N(\bm 0, \hat{\bm\Sigma}^q)$, where $\hat{\bm\Sigma}^q$ is the covariance matrix estimated as described in equation (\ref{cov mat}). At the first look, we generate $\tilde {\bm Z}^1_{(1)}, \tilde {\bm Z}^1_{(2)},\ldots,\tilde {\bm Z}^1_{(V-1)}$ from $N(\bm 0, \hat{\bm\Sigma}^1)$ and obtain maximum test statistic for each $\tilde {\bm Z}^1_{(v)}$ as $\tilde { Z}^1_{\text{MAX}(v)}$ ($v=1,2,\ldots,V-1$). Together with the observed maximum test statistic $Z_{\text{MAX}}^1$, we form the sequence $\mathbb{Z}^1 = \{Z_{\text{MAX}}^1,\tilde {Z}^1_{\text{MAX}(1)},\tilde {Z}^1_{\text{MAX}(2)},\ldots,\tilde {Z}^1_{\text{MAX}(V-1)}\}$ and determine the stopping boundary as the $V(1-\tau_1)$-th smallest value of $\mathbb{Z}^1$, denoted as $b_1 = \mathbb{Q}^{V(1-\tau_1)}(\mathbb{Z}^1)$. At the second look, as each interim analysis is a separate stratum, the incremental information can be incorporated by generating $\tilde {\bm Z}^2_{(1)}, \tilde {\bm Z}^2_{(2)},\ldots,\tilde {\bm Z}^2_{(V-1)}$ from $N(\bm 0, \hat{\bm\Sigma}^2)$. Each maximum test statistic $\tilde {Z}^2_{\text{MAX}(v)}$ is calculated based on $\tilde {\bm Z}^1_{(v)}+\tilde {\bm Z}^2_{(v)}$, or $\bm Z^1+\bm Z^2$ for the observed $Z_{\text{MAX}}^2$. The stopping boundary at the second look is given by $b_2 = \mathbb{Q}^{V(1-\tau_2)}(\mathbb{Z}^2)$. This procedure is repeated at each interim look until either early stopping occurs or the last look is reached. For the choice of $V$, one may follow the recommendation of Finkelstein and Schoenfeld \cite{finkelstein1999combining}, with $V=500$ being sufficient for time-intensive simulations and $V=10,000$ being preferable when feasible.

\section{Simulation Study}
\label{emp com}
In this section, we show the performance of ProFS empirically through simulation. We first compare the power obtained by ProFS and FS, validating ProFS's ability to maintain a specified type I error, and then show the influence of different numbers of examinations in ProFS. For the simplicity of illustration, without loss of generality, we concentrate on the setting with time-to-death and time-to-hospitalization endpoints in our simulation.

\subsection{General Simulation Setup}\label{sim setup}

We consider a two-arm clinical trial with a total sample size of $n=2000$ and equal allocation between the treatment and control groups.
Following Luo et al. \cite{luo2015alternative}, we employ the Gumbel-Hougaard copula with exponential marginal distributions to simulate two correlated times representing the time to death and time to hospitalization endpoints.
Specifically, the vector of time-to-death and time-to-hospitalization in days $(D^*,H^*)$ has the joint survival functions:
$$P(D^*>y_1, H^*>y_2|T) = \exp\left\{-\left[(h_{D}(T)y_1)^\beta + (h_{H}(T)y_2)^\beta\right]^{(1/\beta)}\right\},$$
where $\beta \ge 1$ is the parameter that specifies the correlation between two endpoints, with  Kendall's concordance $W=1 - 1/\beta$. Here, $h_D(T)$ and $h_H(T)$ are treatment-specific hazard rates for death and hospitalization, respectively, and $\beta$ governs the dependence structure between the two endpoints. We consider two values of Kendall’s tau: $W = 0$ (independence) and $W = 0.5$ (moderate positive correlation).

The detailed specifications of $h_{D}(T)$ and $h_{H}(T)$  depend on whether treatment effects are assumed to be constant or primarily short-term. These details will be introduced in the subsequent sections.
We then obtain the observed time to death and time to hospitalization by performing administrative censoring after $S$ days of follow-up, mimicking the limited follow-up window in a real-world clinical trial.
$S_{\text{inf}}$ is set to 0 in the simulation. A significance level of $\alpha = 0.05$ for a two-sided test is applied throughout our simulation.
The empirical power is estimated with 2000 replicates, and the empirical type I error is assessed with 5000 replicates. All computations are implemented in R 4.2.0. An R package 'XX' that implements ProFS will be made publicly available via GitHub upon publication.

\subsection{Performance of ProFS Under Constant Treatment Effects}
\label{section:constant te}
To simulate constant treatment effects, we assume exponential hazards for both endpoints, parameterized by effect size coefficients $\alpha_D$ and $\alpha_H$. Specifically, we let $h_{D}(T) = \lambda_D \exp(-\alpha_D T)$ and $\lambda_{H}(T) = h_H \exp(-\alpha_H T)$ as the hazard rates for death and hospitalization events respectively. We set parameters $\lambda_D = 0.0008, \lambda_H = 0.0022$ and let $\alpha_D,\alpha_H \in \{0,0.1,0.2,0.3\}$ stand for no, very weak, weak, and modest treatment effects, respectively. 

\begin{figure}
 \centerline{\includegraphics[width=6in]{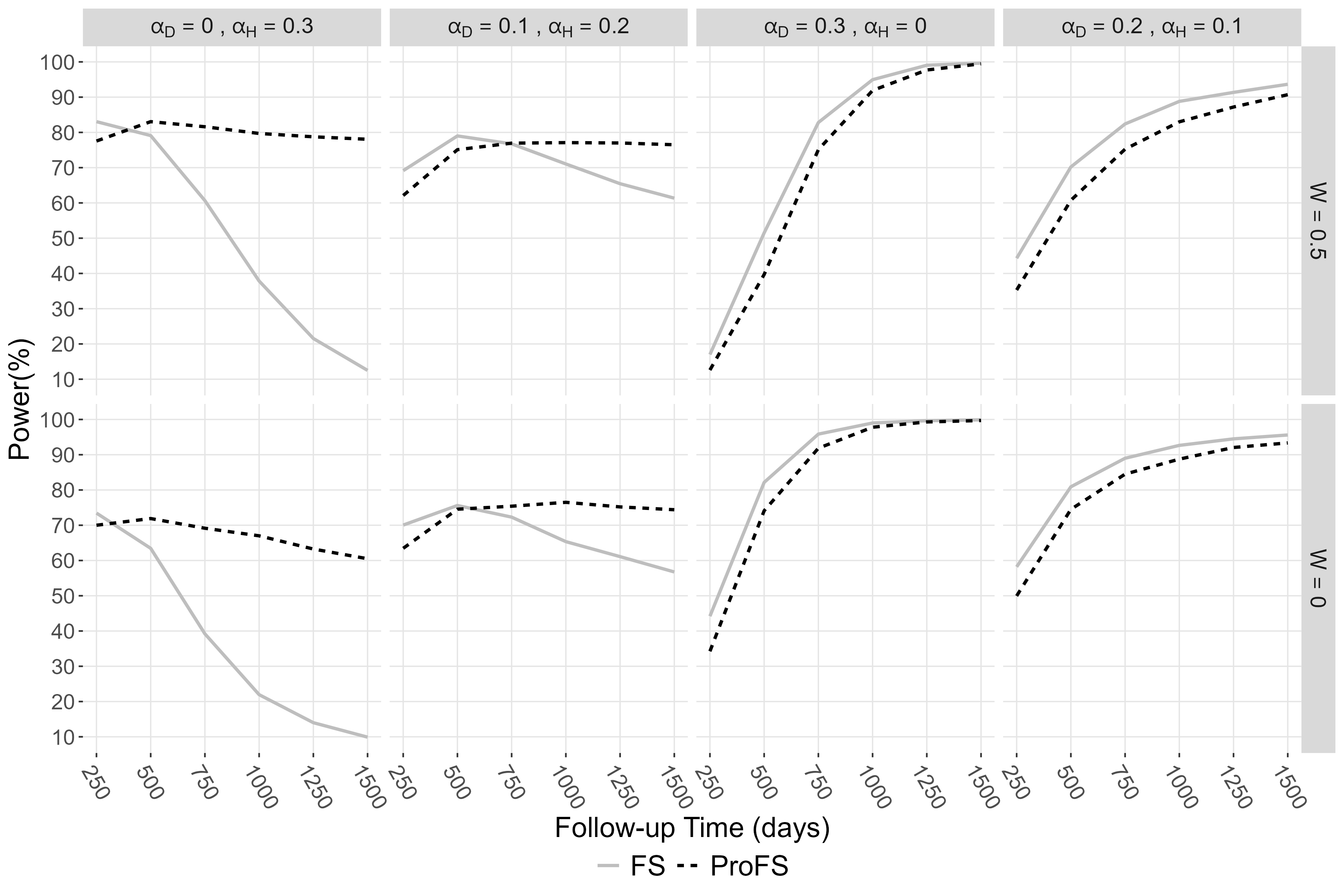}}
\caption{Empirical power of ProFS and FS tests under simulation scenarios with constant treatment effects. $\alpha_D,\alpha_H \in \{0,0.1,0.2,0.3\}$ stand for no, very weak, weak, and modest treatment effects on time-to-death and time-to-hospitalization layers respectively. $W$ stands for Kendall's coefficient of concordance between the two layers.}
\label{fig:sim main}
\end{figure}

The comparative power is presented in Figure \ref{fig:sim main}.  When the treatment effect is limited to the time-to-hospitalization endpoint ($\alpha_D = 0, \alpha_H = 0.3$), FS exhibits a marked decline in power with an extended follow-up period. This contrasts with the scenario where the time-to-hospitalization endpoint is used as a standalone outcome, in which longer follow-up generally yields higher power. In contrast, ProFS sustains a consistent level of power as the follow-up duration increases. As a result, ProFS shows a favorable power for a wide range of follow-up lengths, despite a slightly lower power at the beginning. Here, the lowered power of FS at the longer follow-up time is the result of the hierarchical structure, which gives the time-to-death endpoint higher priority than the time-to-hospitalization endpoint. Since there is no treatment effect on the time-to-death endpoint, more observed death events brought by increased follow-up time make it difficult for FS to detect the true treatment effect on the time-to-hospitalization endpoint \citep{redfors2020win}. We observe a similar pattern when there is a very weak signal for the first layer ($\alpha_D = 0.1, \alpha_H = 0.2$). Although FS shows a temporary increase in power over a short period, it ultimately exhibits a declining trend as the follow-up duration extends further. Overall, ProFS offers favorable robust performance against the choice of follow-up durations compared to FS when there are null or very weak signals in the top layer.   
On the other hand, when there are sufficiently large signals at the top layer ($\alpha_D = 0.2$ or $0.3$), both FS and ProFS exhibit increasing power with longer follow-up durations.
Overall, in such cases, ProFS demonstrates slightly lower power than FS but still delivers comparable performance. This robustness is particularly advantageous when the optimal follow-up length is difficult to determine at the design stage.

Notably, the impact of correlation also differs  depending on whether the top signal is null or very weak ($\alpha_D = 0$ or $0.1$) versus weak or modest ($\alpha_D = 0.2$ or $0.3$). For former, the correlation actually improves the power and the magnitude of improvement is higher for the case with null top layer signal ($\alpha_D = 0$) versus very weak top layer signal ($\alpha_D = 0.1$). Conversely, correlation adversely impacts the power for latter cases with such impact more notable when the top layer has modest signal ($\alpha_D = 0.3,\alpha_H = 0$). Such impacts are caused by the potential spurious negative or positive  ``treatment effects''  observed on the time-to-hospitalization endpoint after conditioning on uninformative comparison on the time-to-death endpoint \citep{verbeeck2019generalized, mou2024winratiomultiplethresholds}. Lastly, when there is no treatment effect ($\alpha_D=\alpha_H=0$), ProFS maintains the empirical type I errors within the acceptable range with Monte Carlo variation under varying follow-up times, as presented in Table \ref{tab:sim examination times} under the column ProFS-4 (ProFS with four examination points).

\subsection{Performance of ProFS Under Short-term Treatment Effects}

\begin{figure}
 \centerline{\includegraphics[width=6in]{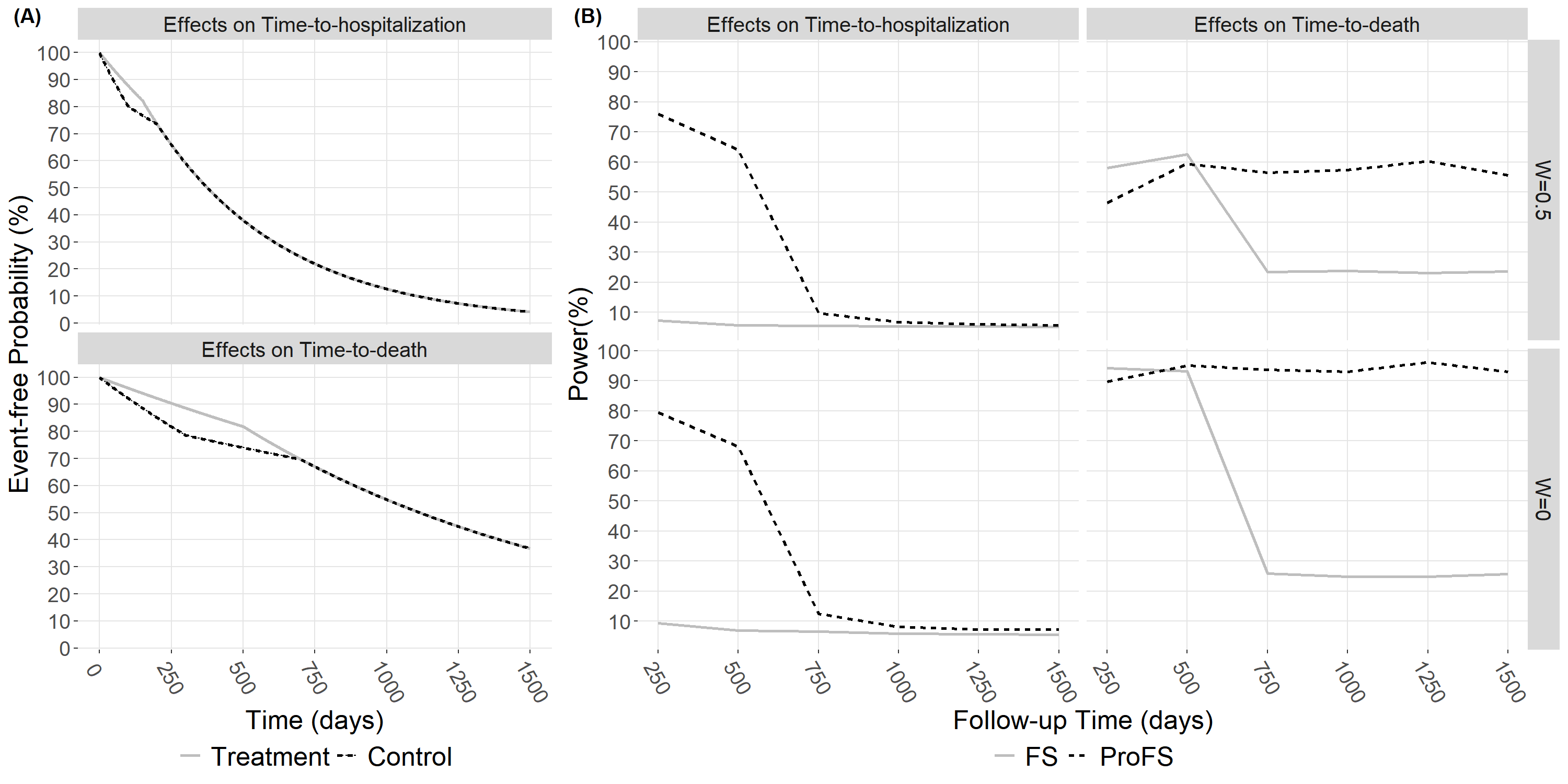}}
 \caption{(A) Theoretical event-free probabilities under short-term treatment effects. Upper subplot: treatment effects are on the time-to-death layer; no treatment effect is on the time-to-hospitalization layer. Lower subplot: treatment effects are on the time-to-hospitalization layer; no treatment effect is on the time-to-death layer. (B) Empirical power of ProFS and FS tests under simulation scenarios with short-term treatment effects on the time-to-death or time-to-hospitalization layers, where $W$ stands for Kendall's coefficient of concordance between two layers.}
\label{fig:sim time varying}
\end{figure}

In this subsection, we consider short-term treatment effects on either the time-to-death or time-to-hospitalization layer. Specifically, when the effect is on the time-to-death layer ($h_H = 0.0022$ for both groups), $h_D$'s  are 0.0004 and 0.0008 for the time intervals $(0, 500]$ and $(500, \infty)$, respectively, for the treatment group, and are 0.0008, 0.0003, and 0.0008 for the time intervals $(0, 300]$, $(300, 700]$, and $(700, \infty)$, respectively, for the control group. When the effect is on the time-to-hospitalization layer ($h_D = 0.0008$ for both groups), $h_H$'s are 0.0013 and 0.0022 for the time intervals $(0, 150]$ and $(150, \infty)$, respectively, for the treatment group, and are 0.00085, 0.0022, and 0.00085 for the time intervals $(0, 100]$, $(100, 200]$, and $(200, \infty)$, respectively, for the control group.  The corresponding event-free curves from those marginal piecewise exponential distributions are depicted in Figure \ref{fig:sim time varying}(A), exhibiting a pattern that conceptually mimics the survival curves as shown in Lederle et al. \cite{lederle2012long}.

The comparative power is shown in Figure \ref{fig:sim time varying}(B). 
When the treatment effect is restricted to the second layer on hospitalization and is short-term, FS consistently experiences a significant lack of power, even when the analysis is confined to a short follow-up period. The deteriorated performance is due to both the hierarchical structure and the dilution of the average treatment effect over the follow-up period when the treatment effect is short-term.  On the other hand, ProFS maintains higher power with a reasonable follow-up length before it begins to decline. This observation confirms that the structure of ProFS enhances the detection of short-term treatment effects. However, its power diminishes with longer follow-up when the earliest examination time stretches to the null effect period. 
When the treatment effect is on the top layer for time-to-death, the FS test starts with favorable power but experiences a sharp decline as the follow-up period extends. In contrast, ProFS sustains a consistent level of power as the follow-up duration increases. As a result, favorable power for a wider range of follow-up time is achieved by ProFS, despite a slightly lower power at the beginning introduced by the penalty for additional examinations.  The impact of the correlation is similar to that observed in the scenarios with constant treatment effects. In summary, the proposed ProFS can be a favorable alternative to FS when the treatment effect is limited to the short term.

\subsection{Number of Examinations in ProFS} \label{subsec_num_exams}

In this subsection, we examine the performance of ProFS with different numbers of examination points under constant treatment effects as specified in Section \ref{section:constant te}. Specifically, ProFS with 2, 4, 5, and 10 quantile examination points, denoted as ProFS-2, ProFS-4, ProFS-5, and ProFS-10, are conducted. The Type I error and empirical power are shown in Table \ref{tab:sim examination times}.
As indicated in Section \ref{section:constant te}, the proposed method is particularly beneficial when signals are primarily in the second layer but may be obscured by a top layer that lacks effect ($\alpha_D = 0, \alpha_H = 0.3$ or $\alpha_D = 0.1, \alpha_H = 0.2$). 
In these simulation scenarios, performance is generally stable across varying numbers of examinations, particularly for four or more.  A notable improvement is observed when increasing from two to four examinations in scenarios with extended follow-up periods, and even more so when there is no signal in the first layer ($\alpha_D = 0$). 
On the other hand, when the signals are primarily in the top layer ($\alpha_D = 0.3, \alpha_H = 0$ or $\alpha_D = 0.2, \alpha_H = 0.1$), increasing the number of examinations may introduce penalties. Nonetheless, performance remains generally stable across varying numbers of examinations.
We also observe that with longer follow-up, the penalty becomes milder. In all cases, the results from four examinations are similar to those from two, especially when compared to the larger number of ten examinations.
Importantly, the empirical type I error rates for different numbers of examinations remain within the acceptable range with Monte Carlo variation, as shown in Table  \ref{tab:sim examination times}.
In summary, ProFS demonstrates reasonable sensitivity to the number of examinations, with the recommended ProFS-4 striking the balance between the benefits of additional examinations and the risks of introducing penalties. Additionally, a larger number of examinations may be a reasonable option when a longer follow-up period is planned.

\begin{table}[htbp]
  \centering
  \caption{Type I error and empirical power (\%) of ProFS with 2, 4, 5, and 10 quantile examinations. The acceptable range of empirical type I error rate with Monte Carlo variation under 5000 replicates is 4.41\% to 5.64\% for the 5\% nominal level.}
      \begin{tabular}{rrrrrrrrr}
    \toprule
   &  $\alpha_D$ & $\alpha_H$ & $W$     & \multicolumn{1}{l}{S} & \multicolumn{1}{l}{ProFS-2} & \multicolumn{1}{l}{ProFS-4} & \multicolumn{1}{l}{ProFS-5} & \multicolumn{1}{l}{ProFS-10} \\
    \midrule
    \multirow{6}{*}{\rotatebox{90}{Type I error}} & 0     & 0     & 0     & 500   & 4.92  & 4.84  & 5.06  & 4.78 \\
    & 0     & 0     & 0     & 1000  & 4.72  & 4.80  & 4.54  & 4.64 \\
    & 0     & 0     & 0     & 1500  & 4.54  & 4.50  & 4.52  & 4.84 \\
    & 0     & 0     & 0.5   & 500   & 5.10  & 4.68  & 4.58  & 4.52 \\
    & 0     & 0     & 0.5   & 1000  & 5.28  & 5.10  & 5.04  & 4.78 \\     
    & 0     & 0     & 0.5   & 1500  & 5.50  & 5.16  & 5.22  & 4.78 \\ \midrule
   \multirow{24}{*}{\rotatebox{90}{Empirical power}} & 0     & 0.3   & 0.5   & 500   & 83.00 & 81.90 & 81.80 & 80.75 \\
    & 0     & 0.3   & 0.5   & 1000  & 73.85 & 79.40 & 79.70 & 79.80 \\
    & 0     & 0.3   & 0.5   & 1500  & 55.10 & 76.10 & 77.10 & 79.30 \\
    & 0     & 0.3   & 0     & 500   & 69.95 & 70.85 & 70.45 & 69.40 \\
    & 0     & 0.3   & 0     & 1000  & 53.85 & 65.20 & 65.45 & 66.75 \\
    & 0     & 0.3   & 0     & 1500  & 31.50 & 58.30 & 62.15 & 65.70 \\
    & 0.1   & 0.2   & 0.5   & 500   & 76.75 & 73.95 & 73.65 & 71.45 \\
    & 0.1   & 0.2   & 0.5   & 1000  & 77.90 & 77.35 & 77.00 & 75.10 \\
    & 0.1   & 0.2   & 0.5   & 1500  & 73.00 & 75.80 & 75.75 & 74.75 \\
    & 0.1   & 0.2   & 0     & 500   & 74.25 & 72.50 & 71.85 & 69.60 \\
    & 0.1   & 0.2   & 0     & 1000  & 71.80 & 73.65 & 73.55 & 72.50 \\
    & 0.1   & 0.2   & 0     & 1500  & 67.40 & 72.00 & 72.10 & 72.70 \\
    & 0.3   & 0     & 0.5   & 500   & 45.30 & 39.95 & 38.45 & 34.25 \\
    & 0.3   & 0     & 0.5   & 1000  & 93.60 & 92.00 & 91.30 & 89.70 \\
    & 0.3   & 0     & 0.5   & 1500  & 99.20 & 99.00 & 98.90 & 98.70 \\
    & 0.3   & 0     & 0     & 500   & 79.55 & 75.20 & 74.25 & 70.55 \\
    & 0.3   & 0     & 0     & 1000  & 98.45 & 97.95 & 97.80 & 97.15 \\
    & 0.3   & 0     & 0     & 1500  & 99.60 & 99.50 & 99.50 & 99.40 \\
    & 0.2   & 0.1   & 0.5   & 500   & 65.55 & 60.25 & 59.55 & 56.15 \\
    & 0.2   & 0.1   & 0.5   & 1000  & 87.90 & 85.25 & 84.35 & 81.20 \\
    & 0.2   & 0.1   & 0.5   & 1500  & 92.70 & 91.60 & 91.25 & 89.10 \\
    & 0.2   & 0.1   & 0     & 500   & 76.85 & 74.45 & 73.05 & 70.00 \\
    & 0.2   & 0.1   & 0     & 1000  & 90.00 & 87.90 & 87.30 & 85.15 \\
    & 0.2   & 0.1   & 0     & 1500  & 93.95 & 93.05 & 92.65 & 91.20 \\
    \bottomrule
    \end{tabular}%
  \label{tab:sim examination times}%
\end{table}%

\section{Case Study}
In this section, we apply the proposed method to analyze the Systolic Blood Pressure Intervention Trial (SPRINT) \cite{sprint2015randomized}. SPRINT was designed to test whether intensive systolic blood pressure control (treatment group) significantly reduces cardiovascular morbidity and mortality compared to the standard treatment (control group) in individuals without diabetes. Of the 14,692 participants, 9,361 were randomized, forming the primary study population. In addition to its primary endpoint, the SPRINT study examined chronic kidney disease (CKD) and related outcomes, where a composite renal endpoint was recorded for participants with baseline CKD. For this case study, we include the primary endpoint and composite renal endpoint as the higher and lower layers in FS and ProFS and demonstrate how ProFS can assist FS in analyzing these outcomes. 
Specifically, the top layer outcome is the primary endpoint, defined as the time to the first occurrence of myocardial infarction (MI), acute coronary syndrome (ACS), stroke, heart failure (HF), or cardiovascular-related death. The second layer outcome is the composite renal endpoint, defined as the time to the first occurrence of end-stage renal disease (ESRD) or a 50\% decline in baseline estimated glomerular filtration rate (eGFR).
We focus on participants with baseline CKD and age $\geq$ 75, as recommended by the SPRINT protocol. This subgroup includes 1,171 participants from 95 clinics. Following the study design, participants were stratified by clinic. Clinics with fewer than five participants were excluded due to their small within-stratum sample sizes, resulting in a final study population of 1,088 participants from 70 clinics.

In the study population, the maximal follow-up time is $S=1704$ days. Since having a sufficient event rate is essential to detect the treatment effect, we require the event rate of the primary event to be at least 10\% at $S_{\text{inf}}$, which leads to a start at $S_{\text{inf}} = 0.58S$ (pooled primary event rate is 10.02\% at $0.58S$). The 4 examination times are $(S_1,S_2,S_3,S_4) = (0.58S, 0.72S, 0.86S, S)$. The test results are presented in Table \ref{tab: cs test result}. Under the significance level of $\alpha=0.05$, ProFS detects a significant difference between the treatment and control groups, while FS concludes no significant difference. This contrasting conclusion appears due to the treatment effect being stronger at $S_2$ and $S_3$ than at $S$, although a formal conclusion on the comparison across different examinations will require further adjustment. 
In summary, ProFS supports the detection of treatment effects and serves as a valuable complement to FS by accounting for the trajectory of increasing follow-up time.

\begin{table}[htbp]
  \centering
  \caption{Hypothesis Testing Results of ProFS and FS Tests}
    \begin{tabular}{lccccc}
    \toprule
          & \multicolumn{4}{c}{ProFS}     & \multicolumn{1}{l}{FS} \\
    \midrule
    Test Statistic & \multicolumn{4}{c}{$Z_{\mathrm{MAX}}$=297} & R=259 \\ 
    p-value & \multicolumn{4}{c}{0.043}     & 0.061 \\
    Examination Time & $S_1$   & $S_2$   & $S_3$   & $S_4$   &  \\
    $R_i$  & 199   & 297   & 262   & 259   &  \\
    \bottomrule
    \end{tabular}%
  \label{tab: cs test result}%
\end{table}%

In this case study, we perform hypothesis tests to assess the presence of a treatment effect and report the corresponding p-values, using a 0.05 significance threshold for illustrative purposes. However, we acknowledge that the results should not be interpreted solely based p-values.
In this post hoc analysis of the SPRINT trial, conducted to demonstrate ProFS, $S_{\text{inf}}$ was determined by identifying the earliest follow-up time at which the event rate reached $10\%$, based on the observed data. While this approach is acceptable for post hoc and secondary analyses, we recommend that, for the primary analysis of a clinical trial, $S_{\text{inf}}$ be determined at the design stage using the prespecified design assumptions, such as the anticipated event rate or hazard rate.

\section{Discussion}

In this study, we propose the ProFS testing method, an extension of FS, to facilitate joint testing of treatment effects across multiple follow-up times, offering advantages in specific scenarios.
Examination times based on quantile values are introduced to simplify their selection in the absence of clinical information. However, the ProFS approach can also align examination times with clinical recommendations when such information is available. 
ProFS thus represents a statistically adaptive and operationally robust generalization that remains anchored in the original prioritization concept while offering enhanced power and interpretability under complex time-to-event structures.
In ProFS, we consider the maximum test for the joint null hypothesis. By incorporating the estimated covariance matrix, this approach considers information from all examinations and serves as a valid global test. An alternative is the global chi-squared test for the multivariate normal distribution. In general, the maximum test tends to be more sensitive to extreme values in tails than the chi-squared test, which may better serve our intended purpose. However, such differences are likely to be small when the number of examinations is not large. A comprehensive comparison between the performance of these two global tests in our context requires further investigation.

There are a few limitations and potential extensions that warrant further
investigation and may inform future methodological developments.
First, extending this concept to endpoints beyond time-to-event endpoints, such as quality-of-life measures, can be challenging unless these measurements are systematically collected and the examination points are appropriately anchored.
The feasibility of employing an imputation model can be investigated, particularly for use in interim analyses. 
For instance,  Broglio et al. \cite{broglio2022comparison} introduced a Bayesian adaptive trial design that includes patients who completed evaluations by an earlier timeline, such as 60 days, with predicted longer-term outcomes incorporated into the interim analysis.
Second, ProFS does not explicitly model the temporal progression of treatment effects. Future work could consider extending ProFS to model longitudinal FS-score processes or incorporate functional representations of treatment effects over continuous time.
Third, currently, ProFS relies on fixed, protocol-specified assessment times. Future enhancements may consider adaptive or data-driven strategies for selecting or aggregating across time points, such as sliding windows or changepoint-informed selection, to better capture dynamic treatment effects. On a related note, while the use of a maximum statistic in ProFS provides strong control of the familywise error rate, it may also lead to conservativeness and reduced sensitivity when the treatment effect is moderate or dispersed over time. Alternative combination methods—such as Simes-type procedures, weighted sums, or threshold-based strategies—could be explored to enhance power while preserving type I error control.
Lastly, the proposed progressive follow-up time framework can be extended beyond FS statistics. Although ProFS is developed to combine FS test statistics for jointly testing treatment effects at multiple time points, the key idea, i.e., including extra examination points and utilizing the maximal test statistic, can be applied to other win statistics. For example, the maximal log win ratio of multiple examinations may be tested in a similar way as long as the joint asymptotic normal distribution of its underlying log win ratios can be obtained. 

\section*{Acknowledgement}
Yunhan Mou's research was supported by CTSA Grant Number UL1 TR001863 from the National Center for Advancing Translational Science (NCATS), a component of the National Institutes of Health (NIH). Its contents are solely the responsibility of the authors and do not necessarily represent the official view of NIH. Dr. Pan's research was supported by the NCI Comprehensive Cancer Center grant (P30 CA021765) and the American Lebanese Syrian Associated Charities (ALSAC). 
We thank the SPRINT study team for making the data available through the Biologic Specimen and Data Repository Information Coordinating Center (BioLINCC) at the National Heart, Lung, and Blood Institute (NHLBI). The authors thank Dr. Vani Shanker for the scientific editing of this manuscript. 

\bibliographystyle{unsrt} 
\bibliography{ref}%

\begin{thebibliography}{10}

\bibitem{pocock2012win}
Stuart~J Pocock, Cono~A Ariti, Timothy~J Collier, and Duolao Wang.
\newblock The win ratio: a new approach to the analysis of composite endpoints
  in clinical trials based on clinical priorities.
\newblock {\em European Heart Journal}, 33(2):176--182, 2012.

\bibitem{finkelstein1999combining}
Dianne~M Finkelstein and David~A Schoenfeld.
\newblock Combining mortality and longitudinal measures in clinical trials.
\newblock {\em Statistics in Medicine}, 18(11):1341--1354, 1999.

\bibitem{FDA2022}
{U.S. Food and Drug Administration (FDA)}.
\newblock Multiple endpoints in clinical trials guidance for industry, 2022.
\newblock Guidance Document.

\bibitem{boentert2024post}
Matthias Boentert, Emmanuelle~Salort Campana, Shahram Attarian, Jordi
  Diaz-Manera, Mazen~M Dimachkie, Magali Periquet, Nathan Thibault, Patrick
  Miossec, Tianyue Zhou, and Kenneth~I Berger.
\newblock Post-hoc nonparametric analysis of forced vital capacity in the comet
  trial demonstrates superiority of avalglucosidase alfa vs alglucosidase alfa.
\newblock {\em Journal of Neuromuscular Diseases}, 11(2):369--374, 2024.

\bibitem{dong2018stratified}
Gaohong Dong, Junshan Qiu, Duolao Wang, and Marc Vandemeulebroecke.
\newblock The stratified win ratio.
\newblock {\em Journal of Biopharmaceutical Statistics}, 28(4):778--796, 2018.

\bibitem{gasparyan2021adjusted}
Samvel~B Gasparyan, Folke Folkvaljon, Olof Bengtsson, Joan Buenconsejo, and
  Gary~G Koch.
\newblock Adjusted win ratio with stratification: calculation methods and
  interpretation.
\newblock {\em Statistical Methods in Medical Research}, 30(2):580--611, 2021.

\bibitem{luo2015alternative}
Xiaodong Luo, Hong Tian, Surya Mohanty, and Wei~Yann Tsai.
\newblock An alternative approach to confidence interval estimation for the win
  ratio statistic.
\newblock {\em Biometrics}, 71(1):139--145, 2015.

\bibitem{bebu2016large}
Ionut Bebu and John~M Lachin.
\newblock Large sample inference for a win ratio analysis of a composite
  outcome based on prioritized components.
\newblock {\em Biostatistics}, 17(1):178--187, 2016.

\bibitem{dong2016generalized}
Gaohong Dong, Di~Li, Steffen Ballerstedt, and Marc Vandemeulebroecke.
\newblock A generalized analytic solution to the win ratio to analyze a
  composite endpoint considering the clinical importance order among
  components.
\newblock {\em Pharmaceutical Statistics}, 15(5):430--437, 2016.

\bibitem{mao2019alternative}
Lu~Mao.
\newblock On the alternative hypotheses for the win ratio.
\newblock {\em Biometrics}, 75(1):347--351, 2019.

\bibitem{oakes2016win}
D~Oakes.
\newblock On the win-ratio statistic in clinical trials with multiple types of
  event.
\newblock {\em Biometrika}, 103(3):742--745, 2016.

\bibitem{dong2020inverse}
Gaohong Dong, Lu~Mao, Bo~Huang, Margaret Gamalo-Siebers, Jiuzhou Wang, GuangLei
  Yu, and David~C Hoaglin.
\newblock The inverse-probability-of-censoring weighting (ipcw) adjusted win
  ratio statistic: an unbiased estimator in the presence of independent
  censoring.
\newblock {\em Journal of Biopharmaceutical Statistics}, 30(5):882--899, 2020.

\bibitem{dong2021adjusting}
Gaohong Dong, Bo~Huang, Duolao Wang, Johan Verbeeck, Jiuzhou Wang, and David~C
  Hoaglin.
\newblock Adjusting win statistics for dependent censoring.
\newblock {\em Pharmaceutical Statistics}, 20(3):440--450, 2021.

\bibitem{mao2024defining}
Lu~Mao.
\newblock Defining estimand for the win ratio: separate the true effect from
  censoring.
\newblock {\em Clinical Trials}, 21(5):584--594, 2024.

\bibitem{buyse2010generalized}
Marc Buyse.
\newblock Generalized pairwise comparisons of prioritized outcomes in the
  two-sample problem.
\newblock {\em Statistics in Medicine}, 29(30):3245--3257, 2010.

\bibitem{luo2017weighted}
Xiaodong Luo, Junshan Qiu, Steven Bai, and Hong Tian.
\newblock Weighted win loss approach for analyzing prioritized outcomes.
\newblock {\em Statistics in Medicine}, 36(15):2452--2465, 2017.

\bibitem{brunner2021win}
Edgar Brunner, Marc Vandemeulebroecke, and Tobias M{\"u}tze.
\newblock Win odds: an adaptation of the win ratio to include ties.
\newblock {\em Statistics in Medicine}, 40(14):3367--3384, 2021.

\bibitem{yang2021eventct}
Song Yang and James Troendle.
\newblock Event-specific win ratios and testing with terminal and non-terminal
  events.
\newblock {\em Clinical Trials}, 18(2):180--187, 2021.

\bibitem{yang2022eventsim}
Song Yang, James Troendle, Daewoo Pak, and Eric Leifer.
\newblock Event-specific win ratios for inference with terminal and
  non-terminal events.
\newblock {\em Statistics in Medicine}, 41(7):1225--1241, 2022.

\bibitem{verbeeck2023generalized}
Johan Verbeeck, Micka{\"e}l De~Backer, Jan Verwerft, Samuel Salvaggio, Marco
  Valgimigli, Pascal Vranckx, Marc Buyse, and Edgar Brunner.
\newblock Generalized pairwise comparisons to assess treatment effects: Jacc
  review topic of the week.
\newblock {\em Journal of the American College of Cardiology},
  82(13):1360--1372, 2023.

\bibitem{peron2018extension}
Julien P{\'e}ron, Marc Buyse, Brice Ozenne, Laurent Roche, and Pascal Roy.
\newblock An extension of generalized pairwise comparisons for prioritized
  outcomes in the presence of censoring.
\newblock {\em Statistical Methods in Medical Research}, 27(4):1230--1239,
  2018.

\bibitem{deltuvaite2023generalized}
Vaiva Deltuvaite-Thomas, Johan Verbeeck, Tomasz Burzykowski, Marc Buyse,
  Christophe Tournigand, Geert Molenberghs, and Olivier Thas.
\newblock Generalized pairwise comparisons for censored data: an overview.
\newblock {\em Biometrical Journal}, 65(2):2100354, 2023.

\bibitem{mao2021class}
Lu~Mao and Tuo Wang.
\newblock A class of proportional win-fractions regression models for composite
  outcomes.
\newblock {\em Biometrics}, 77(4):1265--1275, 2021.

\bibitem{wang2022stratified}
Tuo Wang and Lu~Mao.
\newblock Stratified proportional win-fractions regression analysis.
\newblock {\em Statistics in Medicine}, 41(26):5305--5318, 2022.

\bibitem{song2023win}
James Song, Johan Verbeeck, Bo~Huang, David~C Hoaglin, Margaret Gamalo-Siebers,
  Yodit Seifu, Duolao Wang, Freda Cooner, and Gaohong Dong.
\newblock The win odds: statistical inference and regression.
\newblock {\em Journal of Biopharmaceutical Statistics}, 33(2):140--150, 2023.

\bibitem{jatoi2016time}
Ismail Jatoi, Hanna Bandos, Jong-Hyeon Jeong, William~F Anderson, Edward~H
  Romond, Eleftherios~P Mamounas, and Norman Wolmark.
\newblock Time-varying effects of breast cancer adjuvant systemic therapy.
\newblock {\em Journal of the National Cancer Institute}, 108(1):djv304, 2016.

\bibitem{lederle2009outcomes}
Frank~A Lederle, Julie~A Freischlag, Tassos~C Kyriakides, Frank~T Padberg,
  Jon~S Matsumura, Ted~R Kohler, Peter~H Lin, Jessie~M Jean-Claude, Dolores~F
  Cikrit, Kathleen~M Swanson, et~al.
\newblock Outcomes following endovascular vs open repair of abdominal aortic
  aneurysm: a randomized trial.
\newblock {\em JAMA}, 302(14):1535--1542, 2009.

\bibitem{lederle2012long}
Frank~A Lederle, Julie~A Freischlag, Tassos~C Kyriakides, Jon~S Matsumura,
  Frank~T Padberg~Jr, Ted~R Kohler, Panagiotis Kougias, Jessie~M Jean-Claude,
  Dolores~F Cikrit, and Kathleen~M Swanson.
\newblock Long-term comparison of endovascular and open repair of abdominal
  aortic aneurysm.
\newblock {\em New England Journal of Medicine}, 367:1988--1997, 2012.

\bibitem{lehmann1963robust}
EL~Lehmann.
\newblock Robust estimation in analysis of variance.
\newblock {\em The Annals of Mathematical Statistics}, 34(3):957--966, 1963.

\bibitem{genz1993comparison}
Alan Genz.
\newblock Comparison of methods for the computation of multivariate normal
  probabilities.
\newblock {\em Computing Science and Statistics}, 25:400--405, 1993.

\bibitem{Genz1992numericalcompu}
Alan Genz.
\newblock Numerical computation of multivariate normal probabilities.
\newblock {\em Journal of Computational and Graphical Statistics},
  1(2):141--149, 1992.

\bibitem{genz2009computation}
Alan Genz and Frank Bretz.
\newblock {\em Computation of multivariate normal and t probabilities}.
\newblock Springer Science \& Business Media, Berlin, 2009.

\bibitem{pocock1977group}
Stuart~J Pocock.
\newblock Group sequential methods in the design and analysis of clinical
  trials.
\newblock {\em Biometrika}, 64(2):191--199, 1977.

\bibitem{o1979multiple}
Peter~C O'Brien and Thomas~R Fleming.
\newblock A multiple testing procedure for clinical trials.
\newblock {\em Biometrics}, 35(3):549--556, 1979.

\bibitem{redfors2020win}
Bj{\"o}rn Redfors, John Gregson, Aaron Crowley, Thomas McAndrew, Ori
  Ben-Yehuda, Gregg~W Stone, and Stuart~J Pocock.
\newblock The win ratio approach for composite endpoints: practical guidance
  based on previous experience.
\newblock {\em European Heart Journal}, 41(46):4391--4399, 2020.

\bibitem{verbeeck2019generalized}
Johan Verbeeck, Ernest Spitzer, Ton de~Vries, Gerrit~Anne van Es, WN~Anderson,
  NM~Van~Mieghem, MB~Leon, Geert Molenberghs, and Jan Tijssen.
\newblock Generalized pairwise comparison methods to analyze (non)prioritized
  composite endpoints.
\newblock {\em Statistics in Medicine}, 38(30):5641--5656, 2019.

\bibitem{mou2024winratiomultiplethresholds}
Yunhan Mou, Tassos Kyriakides, Scott Hummel, Fan Li, and Yuan Huang.
\newblock Win ratio with multiple thresholds for composite endpoints, 2024.
\newblock https://arxiv.org/abs/2407.18341.

\bibitem{sprint2015randomized}
{The SPRINT Research Group}.
\newblock A randomized trial of intensive versus standard blood-pressure
  control.
\newblock {\em New England Journal of Medicine}, 373(22):2103--2116, 2015.

\bibitem{broglio2022comparison}
Kristine Broglio, William~J Meurer, Valerie Durkalski, Qi~Pauls, Jason Connor,
  Donald Berry, Roger~J Lewis, Karen~C Johnston, and William~G Barsan.
\newblock Comparison of bayesian vs frequentist adaptive trial design in the
  stroke hyperglycemia insulin network effort trial.
\newblock {\em JAMA Network Open}, 5(5):e2211616--e2211616, 2022.

\end{thebibliography}
\end{document}